\documentclass[sigconf]{acmart}
\AtBeginDocument{%
  }

\usepackage{booktabs}   
\usepackage{multirow}   

\usepackage{booktabs}
\usepackage{siunitx}
\usepackage{balance}
\usepackage{tcolorbox}

\setcopyright{acmlicensed}
\copyrightyear{2026}
\acmYear{2026}
\acmDOI{XXXXXXX.XXXXXXX}
\acmConference[MSR '26]{23rd International Conference on Mining Software Repositories}{April 13--14,
  2026}{Rio de Janeiro, Brazil}

\begin{document}


\title[A Task-Level Evaluation of AI Agents]{A Task-Level Evaluation of AI Agents in Open-Source Projects}

 \author{Shojibur Rahman}

\affiliation{
  \institution{Department of Computer Science, Idaho State University}
  \city{Pocatello}
  \state{Idaho}
  \country{USA}
}
\email{shojiburrahman@isu.edu}

 \author{Md Fazle Rabbi}
 \affiliation{%
   \institution{Department of Computer Science, Idaho State University}
  \city{Pocatello}
  \state{Idaho}
  \country{USA}
}
 \email{mdfazlerabbi@isu.edu}
  
   \author{Minhaz F. Zibran}
 \affiliation{%
   \institution{Department of Computer Science, Idaho State University}
  \city{Pocatello}
  \state{Idaho}
  \country{USA}
}
 \email{zibran@isu.edu }

\renewcommand{\shortauthors}{Rahman et al.}

\begin{abstract}
In this paper, we present a comparative study of five autonomous coding agents using AIDev-pop, which is a public dataset containing thousands of AI-generated pull requests (PRs) across popular open-source repositories. 
We evaluate agents' performance along three task-aware dimensions spanning the PR lifecycle: (1) PR acceptance rate, (2) review discussion volume, and (3) commit message quality. 
Our quantitative analysis finds that Codex consistently achieves high PR acceptance rates across most task categories, while Copilot's PRs trigger the highest volume of both human and automated review discussions. In contrast, commit-level quality varies independently of acceptance outcomes. Claude and Cursor produce higher proportions of high-quality commit messages across several task types, and Codex exhibiting comparatively lower commit quality despite strong integration outcomes. 
Our findings inform selection and improvements of AI agents for their effective integration to collaborative software engineering.
\end{abstract}


\keywords{Quantitative, Analysis, Pull Request, AI Agent, Open-Source, GitHub, AIDev, Empirical Study, Open-Source, Software Engineering
\vspace{0.3cm}
}


\maketitle


\section{Introduction}

Recent advances in large language models have enabled a new class of \emph{autonomous coding agents} that operate beyond interactive code suggestion~\cite{yang2024if}. These agents plan and execute multi-step development tasks, modify multiple files, and submit complete pull requests (PRs) directly to software repositories~\cite{acharya2025agentic}. As a result, AI systems are increasingly act as independent contributors within real-world software engineering workflows.

A growing ecosystem of autonomous coding agents now exists, including OpenAI Codex, Devin, GitHub Copilot, Cursor, and Claude Code~\cite{chen2021evaluating, jimenez2023swe, roychoudhury2025agentic}. Prior work shows that such agents can generate code at scale and contribute to open source development~\cite{chen2021evaluating, jimenez2023swe}. However, our understanding of how different agents compare to one another remains limited, particularly in terms of their technical performance across diverse software engineering tasks~\cite{he2025llm}.
Existing studies examine individual agents or aggregate agent behavior, often emphasizing qualitative observations, productivity gains, or human perceptions of AI-generated contributions~\cite{desolda2025understanding, forsgren2021space, rigby2023improving}. While these perspectives provide valuable insights, they make it difficult to assess relative agent performance under comparable conditions or to identify task-specific strengths and weaknesses~\cite{liang2025swe, jimenez2023swe, he2025llm}. As autonomous coding agents increasingly act as independent contributors in real-world workflows, the lack of task-aware, agent-centric evaluation risks misleading conclusions. A systematic comparison across agents and task types is therefore necessary to support informed tool selection, deployment decisions, and future agent design.

To address this gap, we conduct a comparative, agent-centric study of five widely used autonomous coding agents using the AIDev dataset~\cite{li2025aiteammates_se3}. We analyze objective technical outcomes from real-world PRs and compare agent performance across different PR task types. 
We examine agent behavior across three dimensions: PR acceptance, review interaction, and commit message quality. Specifically, we address the following research questions (RQs):

\vspace{0.2cm}
\label{sec:RQs}
\noindent\textbf{RQ1:} To what extent does PR acceptance differ across autonomous coding agents and PR task types?

\vspace{0.2cm}
\noindent\textbf{RQ2:} To what extent does the volume of review comments differ across agents and PR task types?

\vspace{0.2cm}
\noindent\textbf{RQ3:} To what extent does commit message quality differ across agents and PR task types?

\vspace{0.1cm}
To answer these RQs, we carry out a quantitative analysis using the AIDev~\cite{li2025aiteammates_se3} dataset, which is a recently released large public dataset.
%
%
For transparency and reproducibility, we provide a \emph{replication package}~\cite{ReplicationPackage} publicly available.

\vspace{-0.2cm}
\section{Methodology}
\subsection{Dataset}
The AIDev dataset~\cite{li2025aiteammates_se3} used in this work, is a large collection of PRs authored by five autonomous coding agents (Codex, Devin, Copilot, Cursor, and Claude) in open source GitHub projects, covering data up to August 1, 2025. Specifically, we use the \textit{AIDev-pop}~\cite{aidev_github} subset of the dataset, which includes PRs submitted to repositories at least 100 GitHub stars. This subset contains 33,596 PRs authored by the five autonomous coding agents across 2,807 repositories. This selection allows us to include a broader range of projects. 

Each PR in the AIDev dataset is associated with a task type~\cite{watanabe2025use}, including \texttt{feat} (feature addition), \texttt{fix} (bug fix), \texttt{docs} (documentation), \texttt{build}, \texttt{ci} (continuous integration), \texttt{refactor}, \texttt{test}, \texttt{perf} (performance improvements), \texttt{style}, \texttt{chore} (routine maintenance), \texttt{revert}, and \texttt{other}.
Table~\ref{tab:pr_count_by_task} presents the PR distribution by task type for each AI agent. The \texttt{revert} and \texttt{other} types are rare, with only 16 and 31 PRs across all agents, so we exclude them, leaving 33,549 PRs for analysis.

\vspace{-0.2cm}
\begin{table}[htbp]
\centering
\small
\setlength\tabcolsep{3pt}
\caption{Number of PRs across AI Agents and Task Types}
\label{tab:pr_count_by_task}
\vspace{-0.3cm}
\begin{tabular}{lrrrrrr}
\toprule
\textbf{Task} & \textbf{Claude} & \textbf{Copilot} & \textbf{Cursor} & \textbf{Devin} & \textbf{Codex} & \textbf{Total} \\
\midrule
\textbf{build}    & 8   & 121  & 42  & 64  & 392   & 627   \\
\textbf{chore}    & 14  & 126  & 50  & 281 & 425   & 896   \\
\textbf{ci}       & 5   & 67   & 16  & 59  & 264   & 411   \\
\textbf{docs}     & 32  & 458  & 207 & 620 & 2,570  & 3,887  \\
\textbf{feat}     & 250 & 1,661 & 616 & 1,904 & 10,019 & 14,450 \\
\textbf{fix}      & 115 & 1,993 & 411 & 1,249 & 4,338  & 8,106  \\
\textbf{other}    & 0   & 12   & 9   & 3   & 7     & 31    \\
\textbf{perf}     & 3   & 44   & 24  & 62  & 207   & 340   \\
\textbf{refactor} & 26  & 301  & 111 & 437 & 1,413  & 2,288  \\
\textbf{revert}   & 0   & 5    & 0   & 6   & 5     & 16    \\
\textbf{style}    & 0   & 15   & 18  & 25  & 130   & 188   \\
\textbf{test}     & 6   & 167  & 37  & 117 & 2,029  & 2,356  \\
\midrule
\textbf{Total}    & 459 &4,970 & 1,541 & 4,827 & 21,799 & \textbf{33,596} \\
\bottomrule
\end{tabular}
\vspace{-0.3cm}
\end{table}

\subsection{Criteria for Measurement and Analysis}



\subsubsection{Computation of PR Acceptance Rate:}

To address RQ1, we measure PR acceptance rate as the ratio of merged to submitted PRs for each agent and task type. We compute acceptance rates separately for each agent–task pair to enable fine-grained comparison across software engineering tasks.

\subsubsection{Computing Review Discussion Volume:}

To address RQ2, we quantify discussion volume by measuring the average number of review comments per pull request. We distinguish between comments generated by automated systems (e.g., bots) and those authored by human reviewers. For each agent and task type, we compute the mean number of bot-generated and human-generated comments per PR. This metric captures the degree of review interaction and review overhead associated with agent-generated contributions.

\subsubsection{Assessing Commit Message Quality:}

To assess the quality of a commit messages, as need to address RQ3, we use the \textit{C-Good} classifier proposed by Tian et al.~\cite{tian2022makes}. C-Good is a BERT-based BiLSTM model that identifies a commit message as \textit{high quality} only if it contains both a description of \textit{what} was changed and an explanation of \textit{why} the change was made. Commit messages that miss either the what or the why, or both, are classified as \textit{low quality}. The model was trained and empirically validated through manual labeling and developer surveys, and achieves a precision of 81.6\% in identifying high-quality commit messages~\cite{tian2022makes}.

\section{Analysis and Findings}

\subsection{PR Acceptance Rate of AI Agents (RQ1)}

The acceptance rate for each of the five AI agents across all PR task types is presented in Table~\ref{tab:acceptance_rate}. The results show clear differences in acceptance rates across both agents and task types.

\begin{table}[htbp]
    \centering
    \small
    \setlength\tabcolsep{3pt}
    \caption{PR Acceptance Rate across Agents and Task Types}
\label{tab:acceptance_rate}
\vspace{-0.3cm}
    \begin{tabular}{lrrrrrr}
        \cline{1-6}
        \textbf{Task} & \multicolumn{1}{c}{\textbf{Claude}} & \multicolumn{1}{c}{\textbf{Copilot}} & \multicolumn{1}{c}{\textbf{Cursor}} & \multicolumn{1}{c}{\textbf{Devin}} & \multicolumn{1}{c}{\textbf{Codex}} 
        & \multirow{14}{*}{\rotatebox{90}{SD = Standard Deviation}}
        \\
        \cline{1-6}
        \textbf{build} & \textbf{0.88} & 0.50 & 0.57 & 0.58 & 0.87 \\
        \textbf{chore} & 0.71 & 0.44 & 0.74 & 0.57 & \textbf{0.84} \\
        \textbf{ci} & 0.80 & 0.63 & \textbf{0.94} & 0.61 & 0.86 \\
        \textbf{docs} & 0.75 & 0.61 & 0.74 & 0.71 & \textbf{0.92} \\
        \textbf{feat} & 0.57 & 0.38 & 0.60 & 0.54 & \textbf{0.81} \\
        \textbf{fix} & 0.57 & 0.42 & 0.68 & 0.43 & \textbf{0.82} \\
        \textbf{perf} & 0.67 & 0.27 & 0.46 & 0.35 & \textbf{0.68} \\
        \textbf{refactor} & 0.50 & 0.48 & 0.65 & 0.57 & \textbf{0.80} \\
        \textbf{style} & {-} & 0.40 & 0.72 & 0.68 & \textbf{0.85} \\
        \textbf{test} & 0.50 & 0.37 & 0.62 & 0.50 & \textbf{0.84} \\
        \cline{1-6}
        \textbf{Average} & 0.66 & 0.45 & 0.67 & 0.55 & \textbf{0.83} \\
        \textbf{SD} & 0.13 & 0.11 & 0.13 & 0.11 & \textbf{0.06} \\
        \cline{1-6}
    \end{tabular}
\end{table}

Codex achieves the highest acceptance rates across most task categories. It has the highest overall average acceptance rate (0.83) and the lowest standard deviation (0.06). The low standard deviation shows that Codex’s acceptance rate remains stable across task types. Codex also contributes the largest number of PRs, which shows that its high acceptance rate does not result from a small sample size. In contrast, Copilot records the lowest overall acceptance rate (0.45) and lower acceptance across most task types. Cursor and Claude show moderate average acceptance rates of 0.67 and 0.66, respectively. Both agents exhibit higher variability across task types, with a standard deviation of 0.13.

Acceptance rates differ across task types. Feature additions (\texttt{feat}) and bug fixes (\texttt{fix}) account for the majority of PRs but have lower acceptance rates for most agents. Codex is the only agent with acceptance rates above 0.80 for both task types. Documentation-related PRs (\texttt{docs}) show higher acceptance rates across agents. Codex achieves the highest acceptance rate for documentation tasks (0.92). Maintenance-related tasks such as documentation, build configuration, and continuous integration have higher acceptance rates than functional tasks across agents.

To assess whether the observed differences in acceptance rates are statistically significant, we perform a Mann–Whitney-Wilcoxon (MWW) test~\cite{ahmad1996class} comparing Codex and Cursor, which have the closest average acceptance rates among the agents. The test shows a statistically significant difference in acceptance rates between Codex and Cursor ($p = 4.27 \times 10^{-63}$, $\alpha = 0.05$). This result confirms that the higher acceptance rate observed for Codex is unlikely to be due to random variation. 
Based on the observation and analysis described above, we now derive the answer to RQ1 as follows:

\begin{tcolorbox}[boxrule=0.5pt, boxsep=-2pt, left=5pt, right=5pt]
\textbf{Ans. to RQ1.} PR acceptance differs across both agents and task types. \emph{Codex} consistently achieves \emph{significantly} higher and more stable acceptance rates compared to other agents. PRs for maintenance-related tasks have higher acceptance rates compared to feature additions and bug fixes.
\end{tcolorbox}




\subsection{PR Discussion Volume (RQ2)}
The average number of human-generated and bot-generated comments per PR, aggregated across all task types, is presented in Table~\ref{tab:rq2_discussion}. The results show substantial differences in review comment volume across agents, task types, and reviewer types.

\begin{table}[htbp]
\centering
\small
\setlength\tabcolsep{1pt}
\caption{Average Review Comments per PR across AI Agents, Task Types, and Reviewer Types (i.e., human and bot)}
\label{tab:rq2_discussion}
\vspace{-0.3cm}
\begin{tabular}{@{}lcccccccccc@{}}
\toprule
\textbf{Task} 
& \multicolumn{2}{c}{\textbf{Claude}} 
& \multicolumn{2}{c}{\textbf{Copilot}} 
& \multicolumn{2}{c}{\textbf{Cursor}} 
& \multicolumn{2}{c}{\textbf{Devin}} 
& \multicolumn{2}{c}{\textbf{Codex}} \\
\cmidrule(lr){2-3} \cmidrule(lr){4-5} \cmidrule(lr){6-7} \cmidrule(lr){8-9} \cmidrule(lr){10-11}
\textbf{Type} & Bot & Human & Bot & Human & Bot & Human & Bot & Human & Bot & Human \\
\midrule
\textbf{build}    & 0.00 & 0.00 & \textbf{1.18} & \textbf{1.55} & 0.71 & 0.12 & 0.08 & 0.28 & 0.00 & 0.07 \\
\textbf{chore}    & 0.43 & 0.64 & \textbf{0.66} & \textbf{0.79} & 0.20 & 0.44 & 0.03 & 0.60 & 0.05 & 0.17 \\
\textbf{ci}       & 0.00 & 0.00 & \textbf{0.94} & \textbf{1.16} & 0.00 & 0.50 & 0.14 & 0.76 & 0.01 & 0.02 \\
\textbf{docs}     & 0.00 & 0.41 & \textbf{1.82} & \textbf{1.85} & 0.05 & 0.32 & 0.04 & 1.00 & 0.01 & 0.02 \\
\textbf{feat}     & 0.26 & 0.81 & \textbf{1.74} & \textbf{1.67} & 0.17 & 0.47 & 0.10 & 0.81 & 0.01 & 0.05 \\
\textbf{fix}      & 0.02 & 0.43 & \textbf{0.85} & \textbf{1.04} & 0.04 & 0.14 & 0.05 & 0.32 & 0.01 & 0.04 \\
\textbf{perf}     & 0.00 & 0.00 & \textbf{1.93} & \textbf{1.61} & 0.00 & 0.04 & 0.00 & 0.40 & 0.02 & 0.03 \\
\textbf{refactor} & 0.31 & 0.69 & \textbf{1.40} & \textbf{1.38} & 0.01 & 0.34 & 0.14 & 1.04 & 0.04 & 0.08 \\
\textbf{style}    & --    & --    & \textbf{0.47} & \textbf{0.53} & 0.00 & 0.44 & 0.24 & 0.32 & 0.00 & 0.00 \\
\textbf{test}     & 0.00 & \textbf{1.67} & \textbf{1.49} & 1.50 & 0.05 & 0.11 & 0.31 & 0.44 & 0.01 & 0.02 \\
\hline
\textbf{Average}  & 0.11 & 0.52 & \textbf{1.25} & \textbf{1.31} & 0.12 & 0.29 & 0.11 & 0.60 & 0.02 & 0.05\\
\textbf{SD}       & 0.17 & 0.53 & 0.51 & 0.42 & 0.22 & 0.17 & 0.10 & 0.29 & \textbf{0.02} & \textbf{0.05}\\
\bottomrule
\end{tabular}
\end{table}

Copilot consistently receives the highest number of review comments across nearly all task categories. This pattern holds for both bot-generated and human-generated comments. On average, Copilot PRs receive 1.25 bot comments and 1.31 human comments per PR, which is higher than all other agents. This finding aligns with the acceptance results reported in RQ1, where Copilot exhibits the lowest acceptance rate across agents. Together, these results show that Copilot PRs are associated with higher review effort and lower acceptance outcomes.

In contrast, Codex receives the lowest number of review comments across task types, with averages of 0.02 bot comments and 0.05 human comments per PR, and minimal variation across tasks. A closer examination reveals that the majority of PRs in the dataset receive no review comments at all. Overall, 90.6\% of PRs have zero recorded review comments. This pattern is especially strong for Codex, which contributes the majority of PRs in the dataset. For Codex, 98.2\% of PRs receive no comments. This absence of review comments is especially common for feature additions (\texttt{feat}) and bug fixes (\texttt{fix}), which together account for most Codex PRs. For both task types, 98.2\% of Codex-authored PRs receive no comments. These results show that many PRs, particularly those authored by Codex, are merged or closed with little or no recorded review discussion. As a result, review comment counts capture only explicit written feedback and do not capture all review or decision making activity.

The relationship between human comments and bot comments differs across agents. When bot and human comments are combined, Copilot has an average of 2.56 total comments per PR. All other agents remain below one total comment per PR on average. This indicates that Copilot is the only agent that consistently triggers sustained discussion. It also shows that for most agents, PR review is usually comment free. Human comments dominate bot comments for most agents. For Claude, Devin, Cursor, and Codex, human comments are several times higher than bot comments on average. Copilot is the exception. For Copilot, bot comments and human comments are close in magnitude. This pattern suggests that Copilot PRs trigger both automated feedback and human review, while other agents receive limited feedback from both sources. 

Task type also influences review activity. For Copilot, several tasks show high bot and high human comment volumes, including documentation, feature additions, and performance related tasks. For other agents, comment volume remains low even for tasks where Copilot receives heavy discussion. For example, Codex remains near zero for most tasks for both comment types.

To assess whether differences in review comment volume are statistically significant, we perform a MWW test~\cite{ahmad1996class} comparing Copilot and Devin, which have the closest average review comment volumes among the agents. The test shows a statistically significant difference ($p = 7.09 \times 10^{-50}$, $\alpha = 0.05$), confirming that the higher review comment volume observed for Copilot is unlikely to be due to random variation. 
Based on the observations discussed above, we formulate the answer to RQ2 as follows:


\begin{tcolorbox}[boxrule=0.5pt, boxsep=-2pt, left=5pt, right=5pt]
\textbf{Ans. to RQ2.} Review comment volume varies across agents and task types. \emph{Copilot}'s PRs receive the most bot-generated and human-authored comments, while Codex's PRs receive the fewest. Notably, most PRs, especially those generated by Codex, receive no recorded review comments, which suggests that many PRs are accepted or rejected without documented review discussion.
\end{tcolorbox}




\subsection{Commit Quality (RQ3)}
We apply the \textit{C-Good}~\cite{tian2022makes} classifier to all commits associated with agent-generated pull requests in the dataset, and compute, for each agent and task type, the proportion of commits classified as high quality.
Table~\ref{tab:rq3_commit_quality} presents the proportion of commit messages classified as high quality for each agent across different task types. The results show substantial differences in commit message quality across agents and task types.

\vspace{-0.2cm}
\begin{table}[htbp]
\centering
\small
\setlength\tabcolsep{3pt}
\caption{Good Commit Rate across Agents and Task Types}
\label{tab:rq3_commit_quality}
\vspace{-0.3cm}
\begin{tabular}{lrrrrrr}
\cline{1-6}
\textbf{Task} & \textbf{Claude} & \textbf{Copilot} & \textbf{Cursor} & \textbf{Devin} & \textbf{Codex} & 
\multirow{14}{*}{\rotatebox{90}{SD = Standard Deviation}}
\\
\cline{1-6}
\textbf{build}    & \textbf{0.80} & 0.39 & 0.52 & 0.52 & 0.19 \\
\textbf{chore}    & 0.48 & 0.42 & 0.53 & \textbf{0.56} & 0.34 \\
\textbf{ci}       & 0.40 & 0.46 & \textbf{0.74} & 0.58 & 0.20 \\
\textbf{docs}     & \textbf{0.85} & 0.50 & 0.72 & 0.74 & 0.65 \\
\textbf{feat}     & \textbf{0.70} & 0.34 & 0.52 & 0.57 & 0.19 \\
\textbf{fix}      & \textbf{0.75} & 0.38 & 0.48 & 0.53 & 0.22 \\
\textbf{perf}     & \textbf{1.00} & 0.41 & 0.78 & 0.50 & 0.20 \\
\textbf{refactor} & 0.52 & 0.45 & 0.59 & \textbf{0.63} & 0.49 \\
\textbf{style}    & --    & 0.23 & \textbf{0.68} & 0.53 & 0.28 \\
\textbf{test}     & \textbf{0.77} & 0.52 & 0.76 & 0.62 & 0.41 \\
\cline{1-6}
\textbf{Average}  & \textbf{0.68} & 0.41 & 0.63 & 0.57 & 0.32 \\
\textbf{SD}       & 0.19 & 0.08 & 0.12 & \textbf{0.07} & 0.16 \\
\cline{1-6}
\end{tabular}
\end{table}
\vspace{-0.2cm}

Claude achieves the highest overall good commit rate, with an average of 0.68 across task types. However, Claude also exhibits the highest variability in commit message quality among all agents, with a standard deviation of 0.19. Claude records the highest good commit rates for several task categories, including \texttt{build}, \texttt{docs}, \texttt{feat}, \texttt{fix}, \texttt{perf}, and \texttt{test}. These results show that Claude frequently produces high-quality commit messages, although its performance varies across tasks.

Cursor follows closely, with an average good commit rate of 0.63. Cursor performs particularly well on \texttt{ci} and \texttt{style} tasks, where it achieves the highest good commit rates among agents. Devin shows moderate commit message quality, with an average of 0.57, and exhibits the most stable performance across task types, as reflected by the lowest standard deviation (0.07). Devin performs relatively well on maintenance-oriented tasks such as \texttt{chore} and \texttt{refactor}, where it records the highest good commit rates.

Codex records the lowest overall good commit rate, with an average of 0.32 across task types, and shows relatively high variability (SD = 0.16). The low quality of the commit message for Codex is consistently observed in both functional and maintenance tasks. This finding contrasts with the results of RQ1 and RQ2, where Codex achieves high PR acceptance rates and receives minimal review comments. Given that Codex contributes the majority of PRs in the dataset, these results indicate that many accepted PRs contain commit messages that are not classified as high quality.

To assess whether the observed differences in commit message quality are statistically significant, we perform pairwise comparisons using the MWW test~\cite{ahmad1996class}. We compare Claude with Cursor, which has the closest average good commit rate among the other agents. The test shows a statistically significant difference in commit message quality between Claude and Cursor ($p = 4.55 \times 10^{-31}$, $\alpha = 0.05$). This result confirms that the observed difference in commit message quality is unlikely to be due to random variation.

Commit message quality also varies by task type. Documentation, testing, and performance-related tasks tend to exhibit higher good commit rates across agents. In contrast, feature additions (\texttt{feat}) and bug fixes (\texttt{fix}) show lower good commit rates for most agents. For Codex, commit message quality for \texttt{feat} and \texttt{fix} tasks is particularly low (0.19 and 0.22, respectively), falling below its already low overall average. Despite this, these task types account for the largest number of accepted PRs. This result shows that PR acceptance does not strongly depend on commit message quality. 
Based on the observation and analysis discussed above, we now derive the answer to RQ3 as follows:

\vspace{-0.1cm}
\begin{tcolorbox}[boxrule=0.5pt, boxsep=-2pt, left=5pt, right=5pt]
\textbf{Ans. to RQ3.} Commit message quality differs across agents and task types. \emph{Claude}'s PRs achieves the highest average commit message quality, while Codex's PRs record the lowest, despite its high PR acceptance rate. Commit message quality does not strongly align with PR acceptance.
\end{tcolorbox}




\section{Threats to Validity}

We measure AI agents' performance using their PR acceptance, review comment volume, and commit message quality. These metrics reflect observable outcomes in the PR process but do not fully capture all possible criteria. An accepted PR may still contain quality issues, and review comment volume does not distinguish between positive and negative feedback. 

Factors beyond agent behavior may also affect the results. Task complexity can differ within the same task type, and reviewer practices vary across projects. We reduce these effects by analyzing results at the task level and by using a shared dataset in which all agents operate under comparable repository contexts.

Our analysis focuses on AI-authored PRs in open-source GitHub repositories. The findings may not generalize to proprietary software, other platforms, or settings with close human–AI collaboration. The dataset captures agent behavior at a specific time, and later versions of these agents may behave differently.

\section{Related Work}



Prior work on AI-assisted software development has studied code generation, automated program repair, and AI-based programming support tools~\cite{chen2021evaluating, bouzenia2024repairagent, desolda2025understanding}. These studies often focus on developer productivity and software quality, in interactive settings such as code completion~\cite{forsgren2021space}. More recent research has examined autonomous coding agents capable of performing multi-step development tasks and submitting pull requests to open-source repositories~\cite{hassan2024towards, roychoudhury2025agentic}.

Several empirical studies analyzed AI-generated contributions in real-world development workflows, using metrics such as pull request acceptance and review effort~\cite{measuring-the-impact-of-early-2025-ai-on-experienced-open-source-developer-productivity, rigby2023improving}. The AIDev dataset provides a large-scale view of autonomous AI agents acting as contributors in open-source projects~\cite{li2025aiteammates_se3}. Other work focuses on individual agents or aggregate AI behavior, often emphasizing productivity or qualitative observations~\cite{forsgren2021space, desolda2025understanding, liang2024can}.

Prior research on pull request workflows and code review establishes acceptance rates and review discussion volume as indicators of integration outcomes and review effort~\cite{rigby2023improving}. Separate work on commit message quality highlights the importance of clear documentation for software maintenance and proposes automated assessment methods~\cite{long2003ordinal}.
Our work complements this literature by providing a task-aware, cross-agent evaluation of five autonomous coding agents~\cite{he2025llm}. We compare agent behavior across pull request acceptance, review interaction, and commit message quality using real-world open-source data.



\section{Ethical Considerations}
We analyze publicly available GitHub data from the AIDev dataset and do not involve any private or personal information.

\section{Conclusion}

In this paper, we present a task-aware empirical comparison of five autonomous coding agents using real-world PRs from the AIDev-pop dataset. We analyze agent behavior across PR acceptance, review comment volume, and commit message quality, capturing multiple stages of the PR lifecycle across software engineering tasks. 
Our results show differences across agents, task types, and evaluation dimensions. Codex achieves the highest acceptance rates but produces the lowest-quality commit messages. Copilot triggers the highest review comment volume and has the lowest acceptance rate. Claude produces the highest-quality commit messages with greater task-level variability. These findings show that acceptance, review effort, and commit message quality capture different aspects of agent performance, highlighting the need for task-aware evaluation. Single metrics or aggregate rankings fail to capture these differences, with implications for tool selection and deployment.

We plan to extend this study with a comparison between AI-authored and human-authored pull requests, incorporating code-level quality and defect measures, plus an analysis of how agent behavior changes over time as models evolve. 
%
Our future work will also examine additional signals, such as commit diffs, reviewer identities, or repository-specific practices, to better understand acceptance and rejection outcomes in the absence of explicit review feedback. We will also explore hybrid human–AI workflows to study how autonomous agents integrate into collaborative software development.

\balance
\bibliographystyle{ACM-Reference-Format}
\bibliography{references}

\end{document}